# Investigating Impacts of Health Policies Using Staggered Difference-in-Differences: The Effects of Adoption of an Online Consultation System on Prescribing Patterns of Antibiotics

Kate B. Ellis[1], Ruth H. Keogh[1], Geraldine M. Clarke[2], Stephen O'Neill[3]

[1] Department of Medical Statistics, London School of Hygiene and Tropical Medicine, London, UK

[2] Improvement Analytics Unit, The Health Foundation, London, UK

[3] Department of Health Services Research and Policy, London School of Hygiene and Tropical Medicine, London, UK

*Address for correspondence:* Ruth H. Keogh, Department of Medical Statistics, London School of Hygiene and Tropical Medicine, London, UK, WC1E 7HT, E-mail: Ruth.Keogh@lshtm.ac.uk

## Abstract

We use a recently proposed staggered difference-in-differences approach to investigate effects of adoption of an online consultation system in English general practice on antibiotic prescribing patterns. The target estimand is the average effect for each group of practices (defined by year of adoption) in each year, which we aggregate across all adopting practices, by group, and by time since adoption. We find strong evidence of a positive effect of adoption on antibiotic prescribing rates, though the magnitude of effect is relatively small. As time since adoption increases, the effect size increases, while effects vary across groups.

**Keywords:** antibiotics, difference-in-differences, online consultation system, parallel trends, policy evaluation, staggered adoption

**Running Head:** Effects of Adoption of an Online Consultation System

*We declare no conflicts of interest.*

## 1. Introduction

The use of digital tools in health care settings is growing and has been accelerated by the COVID-19 pandemic. In the United Kingdom (UK) there has been increased adoption of online consultation (OC) systems by general practitioner (GP) practices. These systems enable patients to contact their GP practice by submitting an online form with details of the query (Eccles et al., 2019). After this initial contact, patients may be offered face-to-face appointments, or queries may be dealt with remotely by telephone, video or via online message. There are 31 OC systems that are approved for use by the NHS, varying in design, functionality and how they are implemented (Chappell et al., 2023). In this paper, we examine GP practices using the *askmyGP* OC system (https://askmygp.uk/). Practices using *askmyGP* are encouraged to use the system to facilitate "total digital triage". Under this model, all patient contact to the practice is channelled through an OC system, which patients can do themselves or with assistance from administrative staff, creating a single workflow to support the triage of patient contacts (NHS England, 2020). However, adoption of the *askmyGP* OC system does not always result in total digital triage.

Adoption of an OC system aims to improve the experience of patients by being convenient and accessible, and of practitioners through increasing efficiency of their practice (Eccles et al., 2019). However, it is important to investigate whether adoption could have any unintended implications. Antibiotic prescribing is of particular concern as the misuse and overuse of antibiotics accelerates antibiotic resistance (World Health Organisation, 2020). Since the majority of antibiotics in England are prescribed within general practice (UK Health Security Agency, 2022), this motivates assessing whether adoption of an OC system in general practice causes a change in antibiotic prescribing patterns.

Health policy interventions are often introduced at a cluster level, with roll out being staggered across different groups of units over time. This is the case with OC systems, which have been adopted by different GP practices at different times. When evaluating the effects of adoption, consideration is required as to how to make use of outcomes measured for a GP practice before and after adoption. We focus on a situation in which GP practice-level data is available across a series of time periods, with information in each period on whether a GP practice has adopted the OC system, on the outcome and on a number of characteristics.

When all treated units initiate treatment at the same time, the standard two-way fixed effects difference-in-differences (DiD) estimator is commonly used and provides an unbiased estimate of the average treatment effect in the treated (ATT) provided identification assumptions such as the parallel trends assumption hold (de Chaisemartin and D'Haultfœuille, 2022). However, when roll out of an intervention is staggered over time, recent literature has shown that this estimator can be biased for the ATT when there is treatment effect heterogeneity over time (Goodman-Bacon, 2021, de Chaisemartin and D'Haultfœuille, 2020). The bias arises due to inappropriate comparisons of the outcomes between units that initiate the intervention at different time points (Baker et al., 2022). Several methods have since been proposed to handle variation in timing of treatment initiation (Callaway and Sant'Anna, 2021, Sun and Abraham, 2021, Wooldridge, 2021, Borusyak et al., 2022, de Chaisemartin and D'Haultfœuille, 2020). These methods overcome the issues surrounding standard two-way fixed effect regressions by explicitly avoiding using inappropriate units as controls. In this paper, we consider one such method proposed by Callaway and Sant'Anna (2021) which estimates the average treatment effect for each group of treated units (defined by the time period of treatment initiation) on outcomes measured in each post-treatment time period. These ATTs are referred to as group-time average treatment effects (GTATTs) and the estimation approach is hereby referred to as the staggered DiD approach.

We apply the staggered DiD approach to practice-level data to assess whether adoption of the *askmyGP* OC system caused a change in the antibiotic prescribing habits of health care professionals in English general practice between March 2019 and February 2022. Similar data have previously been used to investigate this topic, but the analysis did not consider the staggered roll out of the system (Dias and Clarke, 2020). The previous evaluation was also restricted to practices that were channelling all patient contact to the practice through *askmyGP* and so investigated the effect of adoption of total digital triage, rather than of the OC system in general. To the best of our knowledge, the staggered DiD approach has not previously been used to investigate effects of adoption of OC systems. However, it been used to assess some implications of the use of telemedicine more broadly. For instance, by using the staggered implementation of Telehealth Parity Laws to assess potential impacts on medical care expenditures in the United States (Dong, 2022).

The remainder of the paper is organised as follows. Section 2 outlines the research question in more detail and describes the data sources. In Section 3, we define the causal estimand of interest and describe the estimation methods, including the standard two-way fixed effects DiD estimator and the staggered DiD approach, in the context of the application. We also discuss methods to assess their identification assumptions. Section 4 describes how the methods are applied to the data and presents the results. The application includes investigations of treatment effect heterogeneity across different dimensions and the sensitivity of the results to certain assumptions.

We conclude with a discussion in Section 5. Additional details on the inclusion criteria, the identification assumptions, and the application results are reported in the online supplementary material accompanying the paper.

## 2. Background to case study and data sources

The aim is to investigate the impact of adoption of the *askmyGP* OC system on antibiotic prescribing rates among GP practices that adopted the system. We carry out our main analyses at the yearly level, making use of practice-level data in each year on whether a GP practice has adopted the OC system, on the outcome and on a number of characteristics. Our analysis period is the 4 full years between March 2018 and February 2022, with $t = 1$ referring to March 2018-February 2019, $t = 2$ referring to March 2019-February 2020 and so on. Assuming any effects of the COVID-19 pandemic prior to March 2020 were negligible in England, we define year periods to run from March to February to avoid straddling the start of the pandemic.

Although our main analyses are carried out at the yearly level between March 2018 and February 2022, we made use of monthly data from May 2017 to March 2022 on use of the system and antibiotic prescribing rates to define our study cohort. All GP practices in England who met certain inclusion criteria over this period were eligible (see Section S.1 of the online supplementary materials for details). We obtained a list of unique identifiers of GP practices that had adopted the *askmyGP* version 3 OC system at any point after its launch in July 2018, up until March 2022. This data included the number of patient-initiated contacts recorded through the system each month for each practice. Our study cohort includes GP practices that had their first patient-initiated contact recorded through the system between March 2019 and February 2021 (during $t = 2$ or $t = 3$), and those that did not adopt the system up to and including February 2022.

The intervention of interest is adoption of the *askmyGP* OC system, which is time-dependent. We define the year of adoption of *askmyGP* as the first year in which there was at least one patient-initiated contact recorded through the system. We assume that the adoption was absorbing, that is, that once a practice first adopted the system, it continued to have access for the rest of our analysis period.

The outcome of interest is the yearly mean antibiotic prescribing rates, where the antibiotic prescribing rates are defined as; the total number of items of antibacterial drugs prescribed each month divided by the monthly practice list size (the number of patients registered to the practice). Antibacterial drugs are classified as such by the British National Formulary (BNF) (OpenPrescribing, 2023a). This data is publicly available from OpenPrescribing, a search interface of the raw English Prescribing Dataset published by the NHS Business Services Authority (OpenPrescribing, 2023b).

We also use data on various GP practice-level characteristics extracted from publicly available data sources (Office for National Statistics, 2022, Office for National Statistics, 2021, GOV.UK, 2019). These sources all produce demographics according to Lower Layer Super Output Area (LSOA), which were then weighted by the proportion of registered patients living in each LSOA to estimate the demographics for patients registered to each eligible GP practice in each month (NHS Digital, 2022). For our main analyses, which are carried out at the yearly level, we use the yearly mean percentage of male patients, percentage of patients of black and minority ethnicity, percentage of patients aged 65 years and over, percentage of patients with third level education, and index of multiple deprivation score of patient area, and a yearly binary indicator for the classification of rurality of patient area.

In total there were 6,397 eligible practices. Of these, 176 (2.75%) adopted the *askmyGP* OC system between March 2019 and February 2021 (during $t=2$ or $t=3$), and we refer to them as the '*askmyGP* practices' throughout. We refer to the remaining practices (who did not adopt the system up to and including February 2022) as the 'never adopters' throughout. Although these practices did not adopt *askmyGP*, they may have been using other OC systems over the analysis period. We therefore assume that their management of patient consultations represents those of the *askmyGP* practices had they not adopted the *askmyGP* system. Of the *askmyGP* practices, 41 adopted the system in the year March 2019-February 2020 ($t=2$), and we refer to them as the 'early adopters'. The remaining 135 *askmyGP* practices adopted during March 2020-February 2021 ($t=3$), and we refer to them as the 'late adopters'. Figure 1 displays the time periods under consideration and the year of adoption of the system for each group of practices (defined by year of adoption).

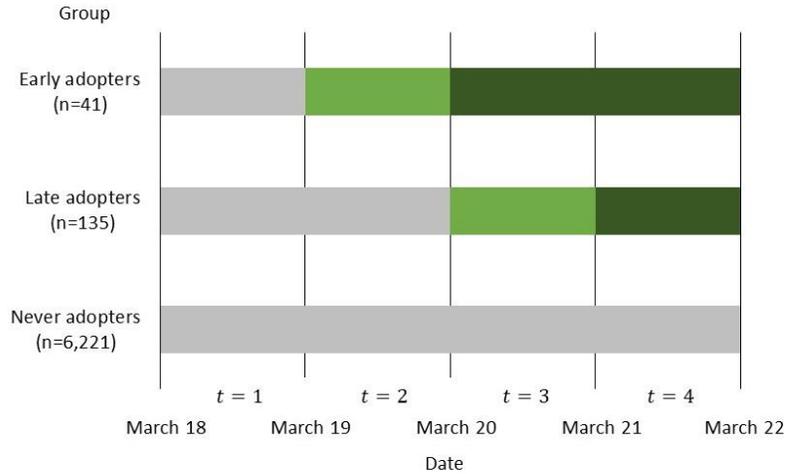

*Figure 1: Illustration of the time periods under consideration and year of adoption of the system for each group of practices (defined by year of adoption). Within each group, years in which no practices have access to the system are highlighted in grey, the year of adoption of the system is highlighted by light green and years in which all practices have access to the system are highlighted in dark green.*

## 3. Methods

### 3.1. Notation and definition of the target estimand

We adopt the notation and potential outcomes framework used by Callaway and Sant'Anna (2021). Let $t = 1, \ldots, T$ denote each time period under consideration. In our application these are years with $t = 1$ referring to March 2018-February 2019 and $T = 4$ referring to March 2021-February 2022. Let $G_{ig}$ denote a binary variable that is equal to 1 if unit $i$ initiates treatment in period $g$, and let $C_i$ denote a binary variable that is equal to 1 if unit $i$ is not treated in any time period under consideration ('never treated'). The observed outcome for unit $i$ in time period $t$ is denoted $Y_{it}$, which in our application is the yearly mean antibiotic prescribing rate. The vector of measured covariates for unit $i$ in time period $t$ is denoted $\boldsymbol{X_{it}}$, which in our application is the vector of yearly practice-level characteristics. Let $Y_{i,t}(0)$ denote the potential outcome for unit $i$ in time period $t$ if they were never treated up to and including time period $T$ and let $Y_{i,t}(g)$ denote the potential outcome for unit $i$ in time period $t$ if they were to have initiated treatment in time period $g$.

The target estimand is the expected difference in the outcome in period $t$ for those who initiate treatment in period $g$ $(g \leq t)$ if all units in this group had initiated treatment in period $g$ compared with if they had never been treated up to and including time period $T$;

$$\text{ATT}(g,t) = E[Y_t(g) - Y_t(0) | G_g = 1] \qquad (1)$$

This is the ATT in time period $t$ for the group of units that initiated treatment in time period $g$. Following Callaway and Sant'Anna (2021), we refer to the target estimand as a group-time average treatment effect (GTATT).

### 3.2. No variation in treatment timing: Two-by-two difference-in-differences

We first consider the simplified setting in which there are only two time periods ($T = 2$) and all units are untreated in the first time period and a group of units are treated in the second. Hence, there are only two groups of units defined by time of treatment initiation in this setting; those that initiate treatment in the second period ($G_{i2} = 1$) and those that are never treated ($G_{i2} = C_i = 0$). Since there are only two time periods and two groups of units, this setting is commonly referred to as a two-by-two DiD setup. Figure 2 illustrates an example of this setup in our application.

*Figure 2: Illustration of a two-by-two DiD setup in our application. Within each group, years in which no practices have access to the system are highlighted in grey and the year of adoption of the system is highlighted by light green.*

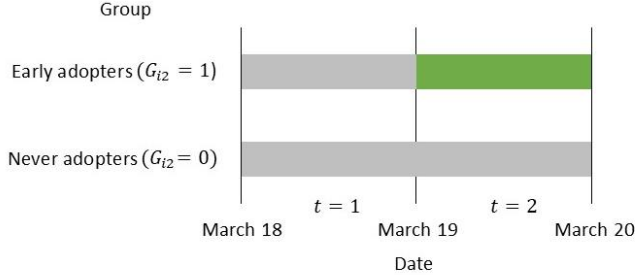

Since there is only one group of treated units, which are observed at one period post-treatment, the target estimand is only defined for $g = 2$ and $t = 2$;

$$\text{ATT} = E[Y_2(2) - Y_2(0) | G_2 = 1] \qquad (2)$$

We consider identification under a *conditional* parallel trends assumption. This assumption states that if the group that initiated treatment in time period $g$ had in fact never been treated, their expected outcomes, conditional on measured covariates, would have followed a parallel path over time to the expected outcomes for those that never became treated. Under this assumption, together with the no anticipation and consistency assumptions (see Section S.2 of the online supplementary materials for details), the conditional ATT (CATT), where the conditioning is on the vectors of covariates in time periods $t = 1$ and $t = 2$, can be identified as follows;

$$CATT = E[Y_2(2) - Y_2(0)|\mathbf{X}_1, \mathbf{X}_2, G_2 = 1] \quad (3)$$

$$= E[Y_2(2)|\mathbf{X}_1, \mathbf{X}_2, G_2 = 1] - (E[Y_2(0) - Y_1(0)|\mathbf{X}_1, \mathbf{X}_2, G_2 = 0] + E[Y_1(0)|\mathbf{X}_1, G_2 = 1]) \quad (4)$$

$$= (E[Y_2|\mathbf{X}_1, \mathbf{X}_2, G_2 = 1] - E[Y_1|\mathbf{X}_1, G_2 = 1]) - (E[Y_2|\mathbf{X}_1, \mathbf{X}_2, G_2 = 0] - E[Y_1|\mathbf{X}_1, G_2 = 0]) \quad (5)$$

Since the CATT is the conditional difference in changes over the two time periods between the two groups, it can be estimated by the coefficient of an interaction term between time period and group indicators in a two-way fixed effects regression model (Baker, 2019);

$$Y_{it} = \omega + \alpha_i + \varphi_t + \delta\tau_t G_{i2} + \boldsymbol{\gamma}^T \mathbf{X}_{it} + \epsilon_{it} \text{ for } i = 1, \ldots, n, t = 1,2 \quad (6)$$

where $\alpha_i$ is the unit fixed effect for unit $i$, $\varphi_t$ is the fixed effect for time $t$, $\tau_t$ is the time period indicator where $\tau_t = 1$ if $t = 2$ and 0 otherwise, and $\epsilon_{it}$ are the residuals that are assumed to be normally distributed with conditional mean zero. The unit fixed effects account for differences between units that are the same over time and the fixed effects for time account for changes over the two time periods that are the same for all units. Assuming the treatment does not affect the measured covariates, inclusion of time-varying covariates controls for measured time-varying differences between groups, where any associated changes in outcome would otherwise be wrongly attributed to the treatment (Zeldow and Hatfield, 2021). Assuming model (6), it can be shown that the CATT in (5) is given by the coefficient of the interaction term between time period and group indicators, $\delta$. When one uses a linear model with no interaction between covariates and treatment group, the conditional effect is equal to the marginal effect. Therefore, under this assumption and the identification assumptions, the ATT in (2) is also given by $\delta$.

In the two-by-two DiD setup one cannot assess the conditional parallel trends assumption since trends in pre-treatment periods cannot be examined. When there is more than one pre-treatment period but still no variation in treatment timing, conventionally the assumption is assessed by testing for differences in pre-treatment trends between the two groups. However, Bilinski and Hatfield (2018) propose exploring the potential impact of a violation on the ATT estimate itself. This can be done by comparing the ATT estimate produced under the conditional parallel trends assumption and an ATT estimate that is produced when a linear trend difference between groups that is extrapolated from pre-treatment periods is allowed (see Section S.3 of the online supplementary materials for details) (Bilinski and Hatfield, 2018).

## 3.3. Variation in treatment timing: The staggered difference-in-differences approach

When there is variation in treatment timing, multiple treated groups can be defined by time of treatment initiation. In our application illustrated in Figure 1, there are 4 time periods ($T = 4$) and the year of adoption of the system varies across the practices. 3 groups of practices are defined by year of adoption; those that never adopted ($C_i = 1$), the early adopters who adopted in the second year ($G_{i2} = 1$) and the late adopters who adopted in the third ($G_{i3} = 1$). When the treatment effect is homogenous across time and group, the GTATTs in (1) will be equal for all $t$ and $g$. In which case, under the same assumptions as in Section 3.2, the overall ATT can be identified by the coefficient for an indicator of having initiated treatment by that period in a two-way fixed effects regression model;

$$Y_{it} = \omega + \alpha_i + \varphi_t + \delta \mathbb{I}(t \geq g \cap G_{ig} = 1) + \boldsymbol{\gamma}^T \boldsymbol{X_{it}} + \epsilon_{it} \text{ for } i = 1, \ldots, n, t = 1, \ldots, T \quad (7)$$

This estimator has been shown to be a weighted average of all possible two-by-two DiD estimates that are produced by comparing outcomes of a group of units whose treatment status changes between two time periods to a group of units whose treatment status does not change (Goodman-Bacon, 2021, Callaway and Sant'Anna, 2022). In some of these comparisons, units that have already been treated are used as comparators for later treated units. When treatment effects vary over time $t$, the change in effects for the earlier treated units renders these comparisons misleading (Baker et al., 2022). Since the standard two-way fixed effects DiD estimator includes these estimates in the weighted average, it thus gives a biased estimate of the overall ATT.

When treatment effects vary over time, an alternative is to estimate the GTATTs for all post-treatment combinations of $g$ and $t$, and then aggregate these into an overall ATT estimate. The GTATTs can be estimated using either outcome regression, inverse probability weighting or doubly robust methods (Callaway and Sant'Anna, 2021). Callaway and Sant'Anna (2021) restrict these methods to the use of pre-treatment covariates only to avoid the risk of involving covariates that have been affected by treatment.

The GTATT estimand in (1) can be identified using outcome regression, by using the result that under identification assumptions it can be expressed as

$$\text{ATT}(g, t)^{OR} = E\left[\frac{G_g}{E[G_g]}\left(Y_t - Y_{g-1} - m_{g,t}(\boldsymbol{X})\right)\right] \quad (8)$$

This relies on estimating the conditional outcome change

$$m_{g,t}(X) = E[Y_t - Y_{g-1}|X_{g-1}, C = 1] \qquad (9)$$

which is the expected change in outcome between periods $g-1$ and $t$ conditional on pre-treatment covariates and on never being treated ($C = 1$).

Alternatively, the GTATT estimand can be identified using inverse probability weighting, since under identification assumptions it can be expressed as

$$\text{ATT}(g,t)^{IPW} = E\left[\left(\frac{G_g}{E[G_g]} - \frac{\frac{P_g(X)C}{1-P_g(X)}}{E\left[\frac{P_g(X)C}{1-P_g(X)}\right]}\right)(Y_t - Y_{g-1})\right] \qquad (10)$$

This relies on estimating the propensity score

$$P_g(X) = P(G_g = 1|X_{g-1}, G_g + C = 1) \qquad (11)$$

which is the probability of having initiated treatment in time period $g$, conditional on pre-treatment covariates and on either having initiated treatment in time period $g$ or never being treated. The $\text{ATT}(g,t)^{IPW}$ is a weighted average of the observed changes in outcome for those that are never treated and those that initiate treatment in period $g$. The weights are defined by the propensity scores where observations from those who are never treated that have similar characteristics to those that initiate treatment in period $g$ are up weighted, ensuring that the covariates of the treatment and comparison group are balanced (Baker, 2019).

Combining these two approaches, the GTATT estimand can be also identified using doubly robust methods, since under identification assumptions it can be expressed in terms of the $\text{ATT}(g,t)^{OR}$, or the $\text{ATT}(g,t)^{IPW}$, and a term with mean zero;

$$\text{ATT}(g,t)^{DR} = E\left[\left(\frac{G_g}{E[G_g]} - \frac{\frac{P_g(X)C}{1-P_g(X)}}{E\left[\frac{P_g(X)C}{1-P_g(X)}\right]}\right)(Y_t - Y_{g-1} - m_{g,t}(X))\right] \qquad (12)$$

If either the propensity score model or the conditional outcome change model is correctly specified, under the conditional parallel trends, no anticipation, consistency and overlap assumptions (see Section S.2 of the online supplementary materials for details), where the conditioning is now on the vector of pre-treatment covariates in time period $g-1$, the GTATT in (1) is given by the $\text{ATT}(g,t)^{DR}$ in (12).

Callaway and Sant'Anna (2021) proposed several overall single summaries of the ATT. We consider the average of the average effects of treatment for each group;

$$\theta_{overall} = \sum_g \theta_{group}(g) P(G = g) \tag{13}$$

This is the average effect of treatment experienced by all units that were ever treated and so it has the same interpretation as the ATT in the two-by-two DiD setup (Callaway and Sant'Anna, 2021).

One can also aggregate the GTATTs to assess whether the treatment has a heterogeneous effect by time of treatment initiation (treatment group) or by time since treatment initiation. The average effect of treatment over time for group $g$ is defined as

$$\theta_{group}(g) = \frac{1}{T - g + 1} \sum_{t=g}^{T} ATT(g, t) \tag{14}$$

The average effect of having initiated treatment $e$ periods ago is defined as

$$\theta_{length}(e) = \sum_g I\{g + e \leq T\} ATT(g, g + e) P(G = g | G + e \leq T) \tag{15}$$

Inference for the GTATTs and aggregate summary measures can be conducted using influence functions (Callaway and Sant'Anna, 2021).

When assessing the conditional parallel trends assumption in this setup, one might consider GTATTs estimated for pre-treatment periods. Assuming there is no anticipation, if there is evidence of an effect in pre-treatment periods, then there is evidence against the assumption. However, for methods that estimate treatment effects according to time since treatment initiation such as the staggered DiD approach, Rambachan and Roth (2023) propose presenting confidence intervals for the average effects of having initiated treatment $e$ periods ago (equation (15)) under different potential violations of the assumption as a sensitivity analysis (see Section S.3 of the online supplementary materials for details) (Rambachan and Roth, 2023).

## 4. Application to *askmyGP* data
### 4.1. Descriptive analyses

Table 1 summarizes the practice-level characteristics in February 2019 for the early and never adopters, and in February 2020 for the late and never adopters. Since the early adopters adopted in March 2019-February 2020 and the late adopters adopted in March 2020-February 2021, these are the last pre-adoption months for each group respectively. In each of the pre-adoption months, the early and late adopters were broadly similar to the never adopters with the following exceptions: the early and late adopters had a smaller proportion of registered patients of black and minority ethnicity, were more likely to be located in a rural area and had a somewhat greater proportion of registered patients aged 65 years and older. Also, in February 2019, the early adopters tended to have larger practice list sizes than the never adopters. We treated all these practice characteristics as potential confounders as we did not believe that distributions of these characteristics could be affected by adoption of the system in the time frame considered, and all could plausibly be associated with antibiotic prescribing rates. It seems plausible that a practice may be more likely to adopt the system if they believed their patients would sufficiently utilise it, which could relate to the age distribution of patients for instance.

*Table 1: Summary of practice-level characteristics comparing the early adopters and never adopters in February 2019, and the late adopters and never adopters in February 2020.*

| Practice characteristic | | February 2019 | | February 2020 | |
| --- | --- | --- | --- | --- | --- |
| | | Never adopters (N=6,221) | Early adopters (N=41) | Never adopters (N=6,221) | Late adopters (N=135) |
| Practice list size | Median | 7799 | 9376 | 7962 | 7310 |
| | (range) | (236, 73453) | (3222, 24523) | (221, 84948) | (2154, 30579) |
| Percentage male | Median | 49.2% | 49.3% | 49.2% | 49.3% |
| | (range) | (45.7, 58.1) | (48.0, 56.2) | (45.4, 57.9) | (47.5, 54.1) |
| Percentage black and minority ethnicity | Median | 7.6% | 3.0% | 7.6% | 3.6% |
| | (range) | (0.5, 90.7) | (1.0, 67.3) | (0.5, 90.5) | (0.9, 64.0) |
| Percentage aged 65 years and over | Median | 18.1% | 21.1% | 18.2% | 19.2% |
| | (range) | (3.0, 47.7) | (4.8, 33.6) | (3.0, 48.4) | (4.9, 35.6) |
| Percentage with third level education | Median | 11.7% | 11.7% | 11.7% | 12.1% |
| | (range) | (6.5, 51.5) | (8.1, 16.9) | (6.5, 51.5) | (9.6, 33.8) |
| Deprivation score of patient areas | Median | 21.3 | 18.4 | 21.3 | 21.7 |
| | (range) | (3.4, 68.7) | (5.4, 49.5) | (3.4, 68.8) | (6.2, 45.8) |
| Practices with the majority of patients living in rural areas | Number (%) | 778 (12.5%) | 12 (29.3%) | 775 (12.5%) | 24 (17.8%) |

Figure 3 illustrates the distributions of yearly mean antibiotic prescribing rates per 1,000 patients in the 4 years between March 2018 and February 2022 for the never, early and late adopters. Over the whole analysis period, the early and late adopters tended to prescribe more items of antibiotics per patient compared to the never adopters. Since this was also the case prior to any adoption of the system, this is likely due to variation in practice and patient characteristics. Overall, trends in antibiotic prescribing rates were similar across the groups.

*Figure 3: Box plots of yearly mean antibiotic prescribing rates per 1,000 patients in $t = 1, 2, 3, 4$ for each group of practices (defined by year of adoption). Within each group, years in which no practices have access to the system are highlighted in grey, the year of adoption of the system is highlighted by light green and years in which all practices have access to the system are highlighted in dark green.*

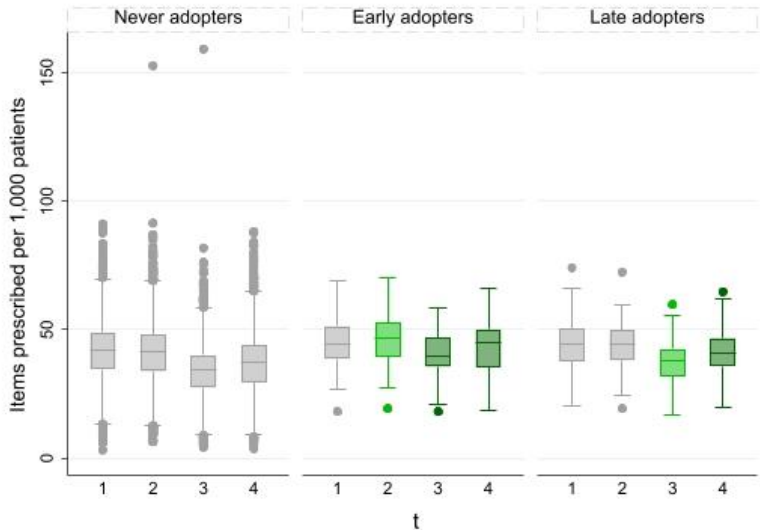

.

## 4.2. Application of the two-by-two difference-in-differences method

We carried out a two-by-two DiD analysis to estimate the effect of adoption of the system for the early adopters in their year of adoption. Figure 2 illustrates the time periods and groups considered for this analysis. We carried out the analysis at the yearly level using the yearly mean antibiotic prescribing rates and practice-level characteristics. Our two-way fixed effects regression model (6) included the following time-varying covariates: yearly mean practice list size, percentage male, percentage of black and minority ethnicity, percentage aged 65 or older, percentage with third level education, deprivation score of patient area, and deprivation score of patient area squared, and the yearly binary indicator for the rural classification. We used cluster–robust standard errors to allow for correlation of prescribing rates over time within practices. To give a better indication of the magnitude of effect, we also estimated the expected percentage increase in average prescribing rates for the early adopters in their year of adoption, compared to if these practices had not adopted it. We estimated this percentage by subtracting the ATT estimate from the observed mean prescribing rate for the early adopters between March 2019 and February 2020 and then divided the ATT estimate by this difference (which is the estimated mean counterfactual for the early adopters between 2019 and February 2020).

The early adopter's average monthly numbers of items of antibiotics prescribed per 1,000 patients between March 2019 and February 2020 were estimated to be 2.3 items higher than if these practices had not adopted the system. There was only very weak evidence of an effect (95% confidence interval (CI)=[-0.4, 4.9], p-value=0.095). This ATT translates into an estimated 5.15% increase in antibiotic prescribing rates for the early adopters in their year of adoption, compared to if these practices had not adopted it.

We carried out Bilinski and Hatfield's (2018) approach to assess the conditional parallel trends assumption for this analysis. Since one cannot use their approach if there is only one pre-treatment period, this was done at the quarterly level using quarterly mean prescribing rates and practice-level characteristics. When assessing the assumption conventionally, one might conclude that there is no evidence against the assumption since there is no evidence of there being a linear trend difference between groups pre-adoption (p-value=0.905). However, when allowing for a linear violation, we cannot rule out changes in the ATT estimate of at least the magnitude of the ATT estimate when assuming no violation. Although this gives some evidence against the conditional parallel trends assumption, we can rule out considerable changes in the ATT estimate when allowing for a linear violation (see Table D1 in the online supplementary materials for details).

### 4.3. Application of the staggered difference-in-differences approach

We applied the doubly robust method of the staggered DiD approach (equation (12)) to the early and late adopters across the 4 years between March 2018 and February 2022, using the never adopters as control units. Again, we carried out the analysis at the yearly level using yearly mean prescribing rates and practice-level characteristics. We used the same covariate specification as in our two-by-two DiD analysis, although now restricted to pre-adoption covariates only. This was done using the *csdid* package in Stata, which is based on their paper (Rios-Avila, 2023). To give a better indication of the magnitude of effect, we also estimated the overall expected percentage increase in average prescribing rates post-adoption of the system, compared to if these practices had not adopted it. We estimated this percentage by subtracting the overall ATT estimate ($\theta_{overall}$, equation (13)) from the observed mean prescribing rate for the *askmyGP* practices between March 2019 and February 2022, and then divided the overall ATT estimate by this difference.

Table 2 gives the estimated GTATTs ($ATT(g,t)$, equation (1)) in each post-adoption year for each group, which all give strong evidence of a positive association between adoption of the system and antibiotic prescribing rates. The estimated overall average effect of adoption of the system experienced by all eligible practices that adopted ($\theta_{overall}$, equation (13)) was 1.7 – this is the expected increase in the monthly numbers of items of antibiotics prescribed per 1,000 patients post-adoption of the system, compared to if these practices had not adopted it. We found strong evidence of there being an effect of adoption when averaged across all those that adopted it (95% CI=[1.1, 2.4], p-value<0.001). This overall ATT translates into an estimated 4.40% increase in antibiotic prescribing rates post-adoption of the system, compared to if these practices had not adopted it.

The GTATT estimates were smaller for the late adopters, compared to the corresponding estimates for the early adopters with the same length of access to the system, suggesting effects of adoption vary across groups. The average effect of adoption was greater for the early adopters ($\theta_{group}(2)$ =2.9 items per 1,000 patients per month, 95% CI=[1.4, 4.4], p-value<0.001), than for the late adopters who had a shorter length of access to the system ($\theta_{group}(3)$ =1.2 items per 1,000 patients per month, 95% CI=[0.6, 1.9], p-value<0.001). Table 3 gives the estimates of the average effect of having adopted 0,1 and 2 years ago ($\theta_{length}(e)$, equation (15)), which suggest that with increasing time since adoption of the system, the effect size increases.

*Table 2: Estimates, 95% confidence intervals and p-values for the GTATTs in periods post adoption (items per 1,000 patients).*

| Group | Time period $t$ | Time periods since adoption $e$ | $ATT(g,t)$ | 95% CI | P-value |
|---|---|---|---|---|---|
| Early adopters ($g = 2$) | 2 | 0 | 2.3 | [0.5, 4.2] | 0.015 |
|  | 3 | 1 | 3.1 | [1.7, 4.5] | <0.001 |
|  | 4 | 2 | 3.2 | [1.3, 5.1] | 0.001 |
| Late adopters ($g = 3$) | 3 | 0 | 0.9 | [0.3, 1.5] | 0.003 |
|  | 4 | 1 | 1.5 | [0.7, 2.3] | <0.001 |

*Table 3: Estimates, 95% confidence intervals and p-values for the average effects of having adopted $e$ periods ago (items per 1,000 patients).*

| Time periods since adoption $e$ | $\theta_{length}(e)$ | 95% CI | P-value |
|---|---|---|---|
| 0 | 1.3 | [0.6, 1.9] | <0.001 |
| 1 | 1.9 | [1.2, 2.6] | <0.001 |
| 2 | 3.2 | [1.3, 5.1] | 0.001 |

Since there was no evidence of an effect of adoption for the late adopters in the year prior to their year of adoption, one might conclude there is no evidence against the conditional parallel trends assumption ($ATT(3,2)$=0.4 items per 1,000 patients per month, 95% CI=[-0.3,1.1], p-value=0.251). However, by carrying out Rambachan and Roth's (2023) approach, we found that our causal conclusions rely quite strongly on the conditional parallel trends assumption holding. For instance, if we were to allow for an exactly linear violation of the conditional parallel trends assumption extrapolated from pre-adoption periods, we would no longer conclude that there is a significant effect of adoption at any period post adoption at the 5% significance level. However, we would still rule out effects of adoption that are large in magnitude (see Table D2 in the online supplementary materials for details).

## 5. Discussion

We used a recently proposed staggered DiD approach (Callaway and Sant'Anna, 2021) to investigate the impact of adoption of an OC system in English general practice on antibiotic prescribing rates, using longitudinal data from 6,397 GP practices. We compared this approach to a more standard DiD method that does not handle the staggered adoption of the system and assessed the validity of our assumptions using recently proposed methods (Bilinski and Hatfield, 2018, Rambachan and Roth, 2023). Our results suggest that adoption of the *askmyGP* OC system increases antibiotic prescribing rates in English general practice, though the magnitude of effect is relatively small. In 2016 NHS England launched a national programme to combat antibiotic resistance where clinical commissioning groups were supported to reduce the number of antibiotics prescribed in primary care by 4% (NHS England, 2016). This suggests that even relatively small changes in antibiotic prescribing rates are considered important and an overall estimated increase of 4.40% post-adoption of the system may be non-trivial.

This study did not incorporate information regarding the reasons for which antibiotics were prescribed at an individual level, and further research is required to understand the reasons for an increase. Prescribing of antibiotics is often necessary and increased rates could reflect higher quality care, rather than a lack of adherence to prescribing guidelines. It has been suggested that if ease of access is improved by adoption of an OC system, the threshold of what patients feel the need to contact their practice about could be lowered, potentially generating new patient demand or uncovering previously unmet need (Salisbury, 2021). If there are higher rates of patient-initiated low acuity demand, this would likely lead to some increase in prescribing rates. Another potential mechanism of the observed association may be if GPs' work pressure in fact increased with adoption of an OC system, since this has been associated with increased prescribing of antibiotics (Allen et al., 2022).

Our results suggest that as the time since adoption increases, the effect size increases. This seems credible as effects of adoption would likely increase as practice staff and patients become more accustomed to the system. It also seems plausible that over time, practices would be more likely to be using *askmyGP* to facilitate total digital triage. This would likely lead to a greater improvement of practice efficiency than if patient queries were not always channelled through the system. Under the same mechanism described above, an increase in practice efficiency could also lead to some increase in prescribing rates. There is also suggestion that effects of adoption vary across groups. The early adopters could be considered to be innovative as they adopted the system prior to the COVID-19 pandemic when there was strong guidance from the NHS that all practices must have access to an OC system (NHS England, 2020). This might also correspond to other behaviours of a practice, such as fully utilising the system as expected leading to a greater effect. However, as more practices in England adopted OC systems over time, those that never adopted the *askmyGP* system would be more similar in terms of their management of patient consultations to the *askmyGP* practices, which may partly explain why the estimated effects of adoption were smaller for the late adopters.

A previous evaluation on the effects of adoption of total digital triage on prescribing rates of antibiotics used the synthetic control method (Dias and Clarke, 2020). While they found a slight increase in average antibiotic prescribing rates, they found no evidence of an effect of adoption of total digital triage during 2019 for 19 English practices that adopted between August 2018 and mid-2019 (ATT = 0.61 items per 1,000 patients per month, 95% CI= [-2.2,4.9], p-value=0.534). Their evaluation was restricted to practices using the *askmyGP* OC system to facilitate total digital triage, and their control pool consisted of practices who went on to adopt total digital triage after the analysis period and so were unlikely to be using other OC systems during the period that they were used as comparators. Their findings are therefore not directly comparable to ours. The synthetic control method relies on different assumptions to the DiD methods considered in this paper. In particular, it does not rely on the conditional parallel trends assumption, but on there being a close pre-treatment match between the characteristics of the treatment group and of a weighted combination of control units (Rehkopf and Basu, 2018). Similarly to DiD methods, extensions have been proposed that handle variation in timing of treatment initiation, such as the generalized synthetic control method (Xu, 2017).

The methods considered in this paper rely on the conditional parallel trends assumption. This assumption is violated if there is a confounder that has either a time-varying effect on the outcome or a time-varying difference between the treatment and control group which has not been appropriately adjusted for (Zeldow and Hatfield, 2021). Using two recently proposed methods (Bilinski and Hatfield, 2018, Rambachan and Roth, 2023) we found some evidence against this assumption. Since the practice-level characteristics used in our analyses were only estimates of the characteristics of registered patients to each practice, there would likely have been some residual confounding. Using model (6) for our two-by-two DiD analysis, we control for measured time-varying differences in characteristics between groups but assume that the characteristics have a constant effect on antibiotic prescribing rates over time. Using the staggered DiD approach, we no longer control for time-varying differences in characteristics between groups but do allow for measured pre-adoption characteristics to have a time-varying effect on antibiotic prescribing rates. Although these methods rely on slightly different assumptions, the ATT estimate from our two-by-two DiD analysis is almost identical to the corresponding GTATT estimate from our staggered DiD analysis (see Section S.4 of the online supplementary materials for an additional comparison). However, the standard errors obtained differ and further assessment of the inference procedures of these methods is required.

Since our control pool included practices operating other OC systems, we estimated the effects of adoption of the *askmyGP* OC system compared to not adopting that particular system. It is difficult to determine how generalisable these results may be to other OC systems since they all vary in functionality. Also, these effects would plausibly be smaller than the general effects of adoption of an OC system compared to not adopting any OC system. Our analyses can be considered as intention to treat (ITT) analyses since we included *askmyGP* practices irrespective of whether they used the system to facilitate total digital triage or potentially stopped using the system. Further research is needed to assess potential impacts of adoption of OC systems and alternative statistical methods, such as the generalized synthetic control method (Xu, 2017), should be applied to assess the validity of our results.

**Data availability**

The data used in this study was generated by combining public domain data and practice-level data on use of the *askmyGP* OC system provided by the Health Foundation. There is a Data Sharing Agreement in place between *askmyGP* and the Health Foundation's Improvement Analytics Unit giving the unit permission to carry out research using the data provided. This data is not personal sensitive data; however, it is commercially sensitive and so cannot be provided by the authors.

**Funding**

KE is funded by the NIHR [NIHR Pre-Doctoral Fellowship Programme (NIHR302010)]. The views expressed are those of the author(s) and not necessarily those of the NIHR or the Department of Health and Social Care. RHK is funded by UK Research and Innovation (Future Leaders Fellowship MR/S017968/1).

# Supplemental Material for "Investigating Impacts of Health Policies Using Staggered Difference-in-Differences: The Effects of Adoption of an Online Consultation System on Prescribing Patterns of Antibiotics"

Kate B. Ellis[1], Ruth H. Keogh[1], Geraldine M. Clarke[2], Stephen O'Neill[3]

[1] Department of Medical Statistics, London School of Hygiene and Tropical Medicine, London, UK
[2] Improvement Analytics Unit, The Health Foundation, London, UK
[3] Department of Health Services Research and Policy, London School of Hygiene and Tropical Medicine, London, UK

*Address for correspondence:* Ruth H. Keogh, Department of Medical Statistics, London School of Hygiene and Tropical Medicine, London, UK, WC1E 7HT, E-mail: Ruth.Keogh@lshtm.ac.uk

## Structure of the document

This document provides supplementary material to the paper "Investigating Impacts of Health Policies Using Staggered Difference-in-Differences: The Effects of Adoption of an Online Consultation System on Prescribing Patterns of Antibiotics". Section S.1 provides the inclusion criteria for GP practices. Section S.2 provides details of the estimation method's identification assumptions. Section S.3 provides further details of the methods used to assess the conditional parallel trends assumption. Section S.4 gives some additional results of the application.

### S.1. Inclusion criteria

GP practices included in the analyses fulfilled the following inclusion criteria:

I. Be an English GP practice.
II. Be active in every month between May 2017 and March 2022.
    *If practices had no registered patients and/or had missing values of the number of items of antibiotics prescribed, we took this as an indicator that the practice was not active in that month.*
III. Have a practice list size of at least 100 patients in every month between May 2017 and March 2022.
IV. Not be another prescribing setting that offers atypical services to standard GP practices.
    *This implicitly excludes walk-in centres, extended access services and community health services.*
V. Not have used *askmyGP* version 3 prior to March 2019.
    *Since some practices may have used askmyGP version 2 prior to the launch of version 3 in July 2018, to try to ensure that we only analysed practices where we could be fairly certain of their date of adoption of any version of askmyGP, we excluded 32 practices that were using askmyGP version 3 in any month prior to March 2019.*
VI. Not have first started using *askmyGP* version 3 after February 2021.
    *Practices that adopted askmyGP late (defined as after February 2021) were very likely to have adopted an alternative OC system before February 2021, as there was strong guidance by the NHS at the start of the COVID-19 pandemic in March 2020 that all practices must have access to an OC system (NHS England, 2020). We did not have access to data on adoption of other OC systems and therefore could not exclude such practices directly. We excluded 6 practices that first started using askmyGP version 3 between March 2021 and February 2022.*

### S.2. Identification assumptions

In this section we give further details of each of the estimation method's identification assumptions. The two-by-two DiD method described in Section 3.2 relies on assumptions 1, 2 and 3. The doubly robust method of the staggered DiD approach described in Section 3.3 relies on assumptions 1, 2, 3 and 4.

We first introduce the *unconditional* parallel trends assumption, which is commonly relied upon in DiD analyses;

$$E[Y_t(0) - Y_{t-1}(0)|G_g = 1] = E[Y_t(0) - Y_{t-1}(0)|C = 1] \text{ for each } g \text{ and } t \in \{2, \ldots, T\} \quad (S1)$$

This assumption states that if the group that initiated treatment in time period $g$ had in fact never been treated, their expected outcomes would have followed a parallel path over time to the expected outcomes for those that never became treated. It is violated if there exists a confounder that has either a time-varying effect on the outcome or a time-varying difference between the treatment and control group (Zeldow and Hatfield, 2021). If one has measured and can appropriately adjust for such confounders, as we do in our analyses, an unbiased estimate of the $\text{ATT}(g, t)$ can be obtained under the slightly weaker *conditional* parallel trends assumption.

**Assumption 1: Conditional Parallel Trends**

$$E[Y_t(0) - Y_{t-1}(0)|\boldsymbol{X}, G_g = 1] = E[Y_t(0) - Y_{t-1}(0)|\boldsymbol{X}, C = 1]$$
$$\text{for each } g \text{ and } t \in \{2, \ldots, T\} \quad (S2)$$

where $\boldsymbol{X} = (\boldsymbol{X}_{t-1}, \boldsymbol{X}_t)$ for the two-by-two DiD method and $\boldsymbol{X} = \boldsymbol{X}_{g-1}$ for the staggered DiD approach. This assumption states that if the group that initiated treatment in time period $g$ had in fact never been treated, their expected outcomes, conditional on measured covariates, would have followed a parallel path over time to the conditional expected outcomes for those that never became treated.

**Assumption 2: No Anticipation**

$$E[Y_t(g)|\boldsymbol{X}, G_g = 1] = E[Y_t(0)|\boldsymbol{X}, G_g = 1] \text{ for each } g \text{ and } t \in \{1, \ldots, T\} \text{ with } t < g \quad (S3)$$

where $\boldsymbol{X} = \boldsymbol{X}_t$ for the two-by-two DiD method and $\boldsymbol{X} = \boldsymbol{X}_{g-1}$ for the staggered DiD approach. This assumption states that for those who initiate treatment in time period $g$, in time periods before $g$, conditional on measured covariates, the expected untreated potential outcomes is equal to the expected potential outcomes for the observed treatment initiation time $g$.

**Assumption 3: Consistency**

$$Y_t(g) = Y_t \text{ if } G_g = 1 \text{ and } Y_t(0) = Y_t \text{ if } C = 1 \text{ for each } g \text{ and } t \in \{1, \ldots, T\} \quad (S4)$$

This assumption states that observed outcomes are equal to the potential outcomes for the observed treatment initiation time, including for those that are never treated, for all time periods.

**Assumption 4: Overlap**

For each $g$, there exists some $\varepsilon > 0$ such that $P(G_g = 1) > \varepsilon$ and
$$p_g(X) = P(G_g = 1 | X_{g-1}, G_g + C = 1) < 1 - \varepsilon \text{ almost surely} \tag{S5}$$

This assumption states that the probability of initiating treatment in time period $g$ is positive, and that the propensity scores are uniformly bounded away from 1.

## S.3. Methods for assessing the conditional parallel trends assumption

In this section we give further details on each of the methods used to assess the conditional parallel trends assumption. Researchers often test for differences in expected outcome trends between treatment and control groups, conditional on measured covariates, in periods prior to treatment and use this to assess the plausibility that trends would be parallel in the absence of treatment. These tests often have low power and as such ATT estimates may be biased by pre-existing trends that are not detected with substantial probability (Roth, 2022).

Bilinski and Hatfield (2018) recommend exploring the potential impact of a conditional parallel trends violation on the ATT estimate itself (Bilinski and Hatfield, 2018). Under their approach, one compares the ATT estimate produced under the conditional parallel trends assumption with an ATT estimate that is produced when a linear trend difference between groups that is extrapolated from pre-treatment periods is allowed. When all treated units start treatment at the same time ($T_0$) and there is more than one pre-treatment period, their approach can be implemented by comparing ATT estimates ($\beta$ and $\beta'$) from two two-way fixed effects regression models;

$$Y_{it} = \omega + \alpha_i + \varphi_t + \sum_{k=T_0}^{T} \beta_k \mathbb{I}(t = k \cap G_{iT_o} = 1) + \gamma^T X_{it} + \varepsilon_{it} \tag{S6}$$

$$Y_{it} = \omega' + \alpha'_i + \varphi'_t + \sum_{k=T_0}^{T} \beta'_k \mathbb{I}(t = k \cap G_{iT_o} = 1) + \theta G_{iT_o} t + \gamma'^T X_{it} + \varepsilon'_{it} \tag{S7}$$

These model specifications produce treatment effect estimates for each post-treatment period ensuring that the linear trend difference ($\theta$ in $(S7)$) is estimated only using data in pre-treatment periods. The ATT from each model is then the simple average of these period effects; $\beta = \frac{1}{T-T_0+1} \sum_{k=T_0}^{T} \beta_k$ and $\beta' = \frac{1}{T-T_0+1} \sum_{k=T_0}^{T} \beta'_k$. They recommend using the confidence interval for the difference in ATTs between the two models ($\beta - \beta'$) to assess the potential impact of the violation.

For methods that estimate treatment effects according to time since treatment initiation such as the staggered DiD approach, Rambachan and Roth (2023) recommend presenting confidence intervals for the average effects of having initiated treatment $e$ periods ago ($\theta_{length}(e)$, equation (15)) under different violations of the assumption as a sensitivity analysis (Rambachan and Roth, 2023). One of their methods extrapolates pre-treatment trend differences to post-treatment periods while restricting the slope of the trend difference to not change by more than some positive constant $M$ across all consecutive periods. They suggest that this restriction would be reasonable when groups could be differentially affected by smoothly evolving trends that would plausibly continue after treatment initiation. Another of their methods allows for violations between consecutive post-treatment periods that are no more than some positive constant $\bar{M}$ times the maximum violation between consecutive pre-treatment periods. They suggest that this restriction would be reasonable when there could be confounding shocks that would plausibly be of a similar magnitude in pre- and post-treatment periods. They recommend reporting the values of $M$ and $\bar{M}$ such that the effect estimates are no longer statistically significant at the 5% level, which they refer to as breakdown values.

## S.4 Additional application results

We now give further details on the results of our application of Bilinski and Hatfield's (2018) approach to assess the conditional parallel trends assumption for our two-by-two DiD analysis, which we carried out at the quarterly level using quarterly mean prescribing rates and practice-level characteristics. Table D1 gives estimates and 95% confidence intervals for the ATT produced under the conditional parallel trends assumption ($\beta$), the ATT produced when a linear trend difference is allowed ($\beta'$), the linear trend difference between groups ($\theta$) and the difference in ATTs ($\beta - \beta'$). The 95% confidence interval for the difference in ATTs includes values greater in magnitude than the ATT estimate produced under the conditional parallel trends assumption. Therefore, when allowing for a linear violation, we cannot rule out changes in the ATT estimate of at least the magnitude of the ATT estimate when assuming no violation of the conditional parallel trends assumption. However, we can rule out considerable changes in the ATT estimate.

*Table D1: Estimates and 95% confidence intervals for the ATT produced under the conditional parallel trends assumption ($\beta$), the ATT produced when a linear trend difference is allowed ($\beta'$), the linear trend difference between groups ($\theta$) and the difference in ATTs ($\beta - \beta'$) (items per 1,000 patients).*

| Parameter | Estimate | 95% CI |
|---|---|---|
| ATT produced under the conditional parallel trends assumption ($\beta$) | 2.4 | [0.1,4.6] |
| ATT that is produced when a linear trend difference is allowed ($\beta'$) | 2.3 | [-0.5,5.1] |
| Linear trend difference between groups ($\theta$) | 0.0 | [-0.5,0.5] |
| Difference in ATTs between the two models ($\beta - \beta'$) | 0.1 | [-2.9,3.2] |

We now give further details on the results of our application of Rambachan and Roth's (2023) approach to assess the conditional parallel trends assumption for our staggered DiD analysis, which was done using the *HonestDiD* package in R (Rambachan, 2022). Table D2 gives the breakdown values of $M$ and $\bar{M}$ and the corresponding 95% confidence intervals at the breakdown values for each of the average effects of having initiated treatment $e$ periods ago. For instance, if we allow for violations of the conditional parallel trends assumption between consecutive post-adoption periods that are no more than 90% of the maximum violation in consecutive pre-adoption periods, we would no longer conclude that there is a significant effect of adoption of the system at two years since adoption at the 5% significance level.

*Table D2: Breakdown values of $M$ and $\bar{M}$, and the corresponding 95% confidence intervals under imposed restrictions at the breakdown values for each the average effects of having initiated treatment $e$ periods ago (items per 1,000 patients).*

|  | $e = 0$ | $e = 1$ | $e = 2$ |
|---|---|---|---|
| Breakdown value of $M$ | 0 | 0 | 0 |
| 95% CI under restriction that slope of extrapolated pre-trend can change by no more than the breakdown value of $M$ | [-0.1,1.8] | [-0.4,2.6] | [-0.8,4.8] |
| Breakdown value of $\bar{M}$ | 1.1 | 0.9 | 0.9 |
| 95% CI under restriction of no more violation than the breakdown value of $\bar{M}$ times the maximum violation in pre-treatment periods | [-0.1,2.5] | [0.0,3.8] | [-0.1,6.6] |

To make an additional comparison between the two-by-two DiD method (applied in Section 4.2) and the doubly robust method of the staggered DiD approach (applied in Section 4.3), we carried out a two-by-two DiD analysis to estimate the effect of adoption of the system for the early adopters in their second year of access to the system. The first time period was again March 2018-February 2019 ($t=1$), where neither the early adopters nor the never adopters had adopted the system. The second time period for this analysis was March 2021-February 2022 ($t=4$), the second full year after the early adopters had adopted the system. The resulting ATT estimate (ATT=3.0 per 1,000 patients, 95% CI=[0.1, 5.8], p-value=0.039) was close to the corresponding GTATT estimate from our staggered DiD analysis ($ATT(2,4)$=3.2 per 1,000 patients, 95% CI=[1.3, 5.1], p-value=0.001, Table 2). However, the standard errors produced under these two approaches differ again.